\documentclass{article}

\usepackage{microtype}
\usepackage{graphicx}
\usepackage{subcaption}
\usepackage{booktabs} 

\usepackage{cancel}
\usepackage{siunitx}
\usepackage{tabularx}
\usepackage{lipsum}
\usepackage{enumitem}

\usepackage{gensymb}

\usepackage{hyperref}

\usepackage[preprint]{icml2026}

\usepackage{amsmath}
\usepackage{amssymb}
\usepackage{mathtools}
\usepackage{amsthm}

\newcommand{\diff}{\mathrm d}

\usepackage{siunitx}
\usepackage[capitalize,noabbrev]{cleveref}

\theoremstyle{plain}

\theoremstyle{definition}

\theoremstyle{remark}

\usepackage[textsize=tiny]{todonotes}
\usepackage[acronym]{glossaries}
\glsdisablehyper

\makeglossaries
\newacronym{mlwp}{MLWP}{Machine learning Weather Prediction}
\newacronym{lam}{LAM}{Limited-Area Modeling}
\newacronym{crps}{CRPS}{Continuous Ranked Probability Score}
\newacronym{cdf}{CDF}{Cumulative Distribution Function}
\newacronym{rmse}{RMSE}{Root Mean Squared Error}
\newacronym{ssr}{SSR}{Spread-Skill Ratio}
\newacronym{cdsi}{CDSI}{Climate Downscaling with Stochastic Interpolants}
\newacronym{hclim}{HCLIM}{Harmonie Climate}
\newacronym{esm}{ESM}{Earth System Model}
\newacronym{ec-earth}{EC-Earth}{European Consortium Earth System Model}
\newacronym{cordex}{CORDEX}{Coordinated Regional Downscaling Experiment}
\newacronym{rcm}{RCM}{Regional Climate Model}
\newacronym{pr}{Pr}{Precipitation}
\newacronym{tas}{TAS}{Surface Air Temperature}
\newacronym{mcc}{MCC}{Matthews Correlation Coefficient }
\newacronym{gcm}{GCM}{General Circulation Model}

\icmltitlerunning{Climate Downscaling with Stochastic Interpolants}

\begin{document}

\twocolumn[
\icmltitle{\acrfull{cdsi}}




\begin{icmlauthorlist}
\icmlauthor{Erik Larsson}{STIMA}
\icmlauthor{Ramon Fuentes-Franco}{SMHI,Bolin,SCIE}
\icmlauthor{Mikhail Ivanov}{SMHI,Bolin}
\icmlauthor{Fredrik Lindsten}{STIMA}
\end{icmlauthorlist}

\icmlaffiliation{STIMA}{Division of Statistics and Machine Learning, Linköping University, Linköping, Sweden}
\icmlaffiliation{SMHI}{Rossby Center, Swedish Meteorological and Hydrological Institute, Norrköping, Sweden}
\icmlaffiliation{Bolin}{Bolin Centre for Climate Research, Stockholm University, Stockholm, Sweden}
\icmlaffiliation{SCIE}{Swedish Centre for Impacts of Climate Extremes, Uppsala University, Uppsala, Sweden}

\icmlcorrespondingauthor{Erik Larsson}{erik.larsson@liu.se}

\icmlkeywords{Downscaling, Machine Learning, Diffusion models, Stochastic Interpolants, Generative Models}

\vskip 0.3in
]



\printAffiliationsAndNotice{\icmlEqualContribution} 

\begin{abstract}
Global climate projections rely on computationally demanding Earth System Models (\acrshort{esm}s), which are typically limited to coarse spatial resolutions due to their high cost. To obtain high-resolution projections for regions of interest, it is common to use Regional Climate Models (\acrshort{rcm}s), which are driven by data produced by \acrshort{esm}s as boundary conditions. While more efficient than running \acrshort{esm}s at fine resolution, \acrshort{rcm}s remain expensive and restrict the size of ensemble simulations.

Inspired by recent advances in probabilistic machine learning for weather and climate, we introduce a data-driven climate downscaling method based on stochastic interpolants. Our approach efficiently transforms coarse \acrshort{esm} output into high-resolution regional climate projections at a fraction of the computational cost of traditional \acrshort{rcm}s. Through extensive validation, we demonstrate that our method generates accurate regional ensembles, enabling both improved uncertainty quantification and broader use of high-resolution climate information.
\end{abstract}

\section{Introduction}\label{intro}
As climate change intensifies, there is a growing need for high-resolution regional climate information to assess local impacts and extreme events. Global climate projections from Earth System Models (\acrshort{esm}s), such as those in CMIP6, typically operate at coarse spatial resolutions (around \SI{100}{\kilo\meter}) due to computational constraints \citep{Jones2023ThermodynamicModificationClimate}. Such coarse grids cannot resolve critical mesoscale processes (e.g., convective storms, complex topography) and thus cannot capture fine-scale variability of climate extremes at scales of 1–\SI{10}{\kilo\meter} \citep{Jones2023ThermodynamicModificationClimate, gmd-copernicus-wang}. This limitation necessitates downscaling to produce regional projections fit for local impact assessments.

The Coordinated Regional Downscaling Experiment (\acrshort{cordex}) addresses this gap by coordinating regional climate modeling at higher resolutions. The European branch, EURO-\acrshort{cordex}, has produced ensembles of regional climate simulations at 0.1\degree (around \SI{12}{\kilo\meter}) resolution. These high-resolution simulations are widely used by scientists and stakeholders, forming a critical basis for impact assessments and adaptation planning. Finer spatial detail is essential for simulating climate extremes; for example, high-resolution precipitation data are needed to assess flood risks \citep{gmd-copernicus-wang}. By resolving topography and coastlines, \acrshort{rcm} downscaling can reduce biases in regional precipitation and temperature \citep{gmd-copernicus-wang}. Importantly, \acrshort{rcm}s produce physically consistent fields at fine scales \citep{Jones2023ThermodynamicModificationClimate}, which is crucial for analyzing multi-variable extreme events coherently.

Despite these benefits, \acrshort{rcm}s are computationally intensive, limiting the number of scenarios or ensemble members that can be downscaled \citep{Jones2023ThermodynamicModificationClimate}. A single \acrshort{rcm} run spanning decades at 0.1\degree requires orders of magnitude more computing resources than one at \SI{50}{\kilo\meter} \citep{gmd-copernicus-wang}.

\paragraph{Our main contributions are:}\begin{enumerate}
    \item We propose \acrfull{cdsi}, a multivariate probabilistic data-driven downscaling framework specifically developed to downscale global low-resolution climate simulations to high-resolution regional climate simulations.
    \item We show that \acrshort{cdsi} can directly downscale \acrshort{esm} outputs to high-resolution regional projections, producing ensembles with performance competitive to state-of-the-art diffusion-based approaches while avoiding the multi-model, sequential pipelines used by competitive baseline methods.
    \item We demonstrate that \acrshort{cdsi} can generalize to unseen future conditions and across new climate model realizations, indicating robustness under distribution shift.
\end{enumerate}

\section{Background}
In this study, we use the HARMONIE-Climate (\acrshort{hclim}) regional climate modeling system as our reference dataset, serving as both the ground truth for validation and the training target for our emulator.  \acrshort{hclim} is a community-developed, high-resolution modeling framework designed for regional climate simulations across a wide range of spatial scales. Built upon the HIRLAM–ALADIN Numerial Weather Prediction system, it integrates advanced physical parameterizations, data assimilation techniques, and surface modeling via the SURFEX land model. The system includes multiple configurations optimized for different resolutions: AROME for convection-permitting ($\sim$\SI{2.5}{\kilo\meter}), ALARO for intermediate ($\sim$\qtyrange[range-units=single, range-phrase=-]{4}{10}{\kilo\meter}), and ALADIN for coarser ($\sim$\qtyrange[range-units=single, range-phrase=-]{10}{50}{\kilo\meter}) grids. In this work, we employ simulations from the HCLIM38 configuration using ALADIN physics at \SI{12}{\kilo\meter} resolution over the EURO-\acrshort{cordex} domain. \acrshort{hclim} has been extensively validated and applied in coordinated international projects such as EURO-\acrshort{cordex} and \acrshort{cordex} FPS, and is recognized for its ability to simulate temperature and precipitation variability across Europe with high spatial fidelity. Given its robustness, process realism, and widespread adoption in the regional modeling community, \acrshort{hclim} provides an ideal benchmark for evaluating downscaling approaches like \acrshort{cdsi}.

\paragraph{Problem Definition.}
In this work, we address the problem of downscaling, where the goal is to recover a high-quality (HQ) field from a corresponding low-quality (LQ) input. Specifically, we aim to learn a mapping from LQ to HQ that can enhance spatial resolution and correct systematic errors or biases present in the low-quality data. In addition to LQ, auxiliary static features such as land–sea masks and topography, as well as precomputed dynamic features like time of day or year, may be incorporated to provide additional information to the model.

\paragraph{Downscaling vs. Simulation.}
An alternative approach to obtaining HQ fields from LQ inputs is to perform regional simulation \citep{larsson2025diffusionlam, oskarsson2024probabilistic,Regionaldatadrivenweathermodelingwithaglobalstretchedgrid,pathak2024kilometerscaleconvectionallowingmodel,building_lams, crps_lam}. For regional simulations, the task differs from downscaling. Instead of directly learning a mapping from LQ to HQ, the HQ fields are generated through simulation, where the model evolves trajectories within the HQ domain driven by the LQ dynamics and previous HQ states. This approach offers the advantage of producing temporally consistent trajectories, where each HQ ensemble member naturally evolves from its predecessor. However, it also involves the computational cost of simulating full trajectories, making it impractical for long-term applications such as climate prediction, where simulations may span on the order of a hundred years. In addition, maintaining stable and physically plausible model behavior over long roll-outs can be challenging for data-driven regional simulation approaches, as small errors may accumulate over time. This approach is also impractical when the objective is to generate high-resolution outputs at specific time instances, as it requires unnecessary simulation of the entire temporal evolution.

In contrast, downscaling allows HQ fields to be inferred directly from LQ inputs at any desired time point without requiring the simulation of full trajectories. For this reason, we focus on downscaling as it provides greater flexibility and computational efficiency, particularly for large-scale or long-horizon climate applications.

\paragraph{Downscaling vs. Super-resolution.}
Superresolution is a well-studied problem in computer vision, where the objective typically is to recover a high-resolution image from a low-resolution observation, typically by inverting a known or implicit degradation such as blurring or subsampling \citep{GENDY2025128911}.

Downscaling in geophysical and climate applications differs fundamentally from classical super-resolution. The low-quality input is not obtained by a simple or known degradation operator acting on the high-quality field. Instead, it often originates from a different numerical model, resolution, or parameterization, and therefore exhibits systematic biases, structural errors, and mismatches in physical variability. As a result, downscaling entails not only increasing spatial resolution but also correcting model-dependent biases and inconsistencies. Furthermore, when multiple variables are downscaled simultaneously, the model must learn complex cross-variable dependencies while respecting the differing spatial scales, spectra, and statistical properties of each field, making downscaling a substantially more challenging problem than standard image super-resolution. 

Consequently, downscaling cannot be treated as a straightforward extension of image super-resolution, but rather as a joint resolution enhancement and model correction problem in a multivariate physical system.

\subsection{Related Works}
In recent years, numerous works have explored probabilistic machine learning approaches for weather \citep{springenberg2025diffscalecontinuousdownscalingbias, CorrDiff, watt2024generativediffusionbaseddownscalingclimate, Tomasi_LDM_downscaling,ramon_fuentes} and climate downscaling \citep{Dynamical_generative_downscaling_of_climate_model_ensembles, brenowitz2025climatebottlegenerativefoundation}. However, the lack of standardized benchmarks has led to substantial variation across studies, with different datasets, experimental setups, and downscaling ratios between LQ inputs and HQ outputs. This heterogeneity makes direct comparison between methods challenging and complicates the assessment of relative performance and generalization capabilities.

Previous work has explored the use of diffusion models for climate downscaling. For example, R2-D2 employs a hybrid approach where the \acrshort{esm} is first downscaled to \SI{45}{\kilo\meter} resolution using a \acrshort{rcm} from an initial \SI{100}{\kilo\meter} grid, and then a diffusion model is applied to further refine the resolution to \SI{9}{\kilo\meter}, generating ensemble members \citep{Dynamical_generative_downscaling_of_climate_model_ensembles}. In contrast, our approach directly learns high-resolution regional fields from coarse \acrshort{esm} data, simplifying the workflow and reducing reliance on intermediate simulations with a RCM.

CorrDiff \citep{CorrDiff} is trained to predict \SI{2}{\kilo\meter}-resolution regional weather fields conditioned on \SI{25}{\kilo\meter}-resolution global weather data. Motivated by the observation that diffusion models can struggle to directly model the conditional distribution \( p(\mathrm{HQ} \mid \mathrm{LQ}) \), CorrDiff decomposes the problem into two stages. First, a deterministic model is trained to predict the conditional mean \( p(\mathrm{HQ}_{\mu} \mid \mathrm{LQ}) \). A diffusion model is then used to generate stochastic residuals around this mean, thereby producing ensemble members. While this residual-corrective formulation alleviates some of the challenges associated with directly applying diffusion models to conditional downscaling, it introduces additional complexity by requiring a separate deterministic model. 

\subsection{Diffusion Models}
In this work, we focus on a particular formulation of the diffusion model, the EDM framework proposed by \citet{karras2022elucidating}, as it is used in the EDM and CorrDiff baselines.

The generative process starts from an initial noisy sample $Z_0$ and progressively transforms it into a clean sample $Z_N$ by following the probability flow ordinary differential equation (ODE),
$$
\mathrm{d}x = -\dot{\sigma}(t)\,\sigma(t)\,\nabla_x \log p(x;\sigma(t))\,\mathrm{d}t.
$$
In practice, this ODE is solved numerically using a finite sequence of steps. Denoting the solver update by $D_\theta$, the diffusion trajectory is given by
$$
Z_{n+1} = D_\theta(Z_n, C, \sigma_{n+1}, \sigma_n),
\quad n = 0,1,\dots,N-1,
$$
where the noise level decreases from $\sigma_n$ to $\sigma_{n+1} < \sigma_n$, and the process is conditioned on auxiliary input $C$.

Following the EDM formulation, the solver update $D_\theta$ is parameterized using a neural network $F_\theta$ with noise-dependent preconditioning,
    \begin{multline}
        D_{\theta}(Z_n, C, \sigma_{n+1}, \sigma_n) = 
        c_{\text{skip}}(\sigma_n) \cdot Z_n \\ + c_{\text{out}}(\sigma_n) \cdot F_{\theta}(c_{\text{in}}(\sigma_n) \cdot Z_n, c_{\text{noise}} (\sigma_n), C),       
    \end{multline}
where the preconditioning coefficients are defined as
\begin{align*}
    c_{\text{skip}}(\sigma_n) &= \frac{\sigma_{\text{data}}^2}{\sigma_n^2 + \sigma_{\text{data}}^2}, &
    &c_{\text{out}}(\sigma_n) = \frac{\sigma_n^2\,\sigma_{\text{data}}^2}{\sqrt{\sigma_n^2 + \sigma_{\text{data}}^2}}, \\
    c_{\text{in}}(\sigma_n) &= \frac{1}{\sqrt{\sigma_n^2 + \sigma_{\text{data}}^2}}, &
    &c_{\text{noise}}(\sigma_n) = \tfrac{1}{4}\log(\sigma_n).    
\end{align*}
Since the targets are normalized, we set $\sigma_{\text{data}} = 1$. 

The noise schedule is constructed using a $\rho$-parameterized interpolation,
$$
\sigma_n =
\left(
\sigma_{\text{max}}^{1/\rho}
+ \frac{n}{N-1}
\left(
\sigma_{\text{min}}^{1/\rho}
- \sigma_{\text{max}}^{1/\rho}
\right)
\right)^{\rho},
\qquad \sigma_N = 0.
$$

During training, Gaussian noise $\mathcal{N}(0, \sigma_n^2 I)$ is added to the ground-truth target using a noise level $\sigma_n$, where $n$ is sampled uniformly. A single denoising step is then performed to produce a prediction $\hat{X}$ of the target $X$. The model is trained using a weighted mean squared error loss,
\begin{equation}\label{eq:training_loss}
\mathcal{L}
=
\mathbb{E}_{n \sim \text{Uniform}(0, N-1)}
\left[
\omega_n \, (\hat{X} - X)^2
\right],
\end{equation}
where the loss is averaged over spatial grid points and summed across all predicted variables. Following \citet{karras2022elucidating}, the loss is reweighted according to
$$
\omega_n = \frac{\sigma_n^2 + \sigma_{\text{data}}^2}{(\sigma_n\,\sigma_{\text{data}})^2}
$$
to accounts for the noise level used during training. This weighting reduces the influence of early diffusion steps, where reconstruction errors are naturally larger, and places greater emphasis on later denoising stages that are closer to the data manifold.

For sampling, we use the second-order EDM ODE solver with $N=20$ solver steps. Due to the second-order scheme, this corresponds to $2N-1 = 39$ sequential function evaluations of $D_\theta$ per generated forecast. Ensemble forecasts are obtained by drawing independent initial noise samples $Z_0^{(i)} \sim \mathcal{N}(0, \sigma_0^2 I)$ for each ensemble member.

The hyperparameters used during training and sampling are summarized in \cref{tab:hyperparams}. When sampling, we change the hyperparameters slightly following \cite{karras2022elucidating,gencast} and use a second-order ODE solver as proposed in \citet{karras2022elucidating}.

\begin{table}[h]
\centering
\caption{EDM hyperparameters}
\label{tab:hyperparams}
\begin{tabular}{ccc}
\bf Hyperparameter & \bf Value Training & \bf Value Sampling 
\\ \hline
$\sigma_{\text{max}}$ & $88$ & $80$\\
$\sigma_{\text{min}}$ & $0.02$ & $0.03$\\
$\rho$ & $7$ & $7$
\end{tabular}
\end{table}

\begin{figure*}[t]
    \centering
    \begin{minipage}{\linewidth}
        \centering
        \includegraphics[width=\linewidth]{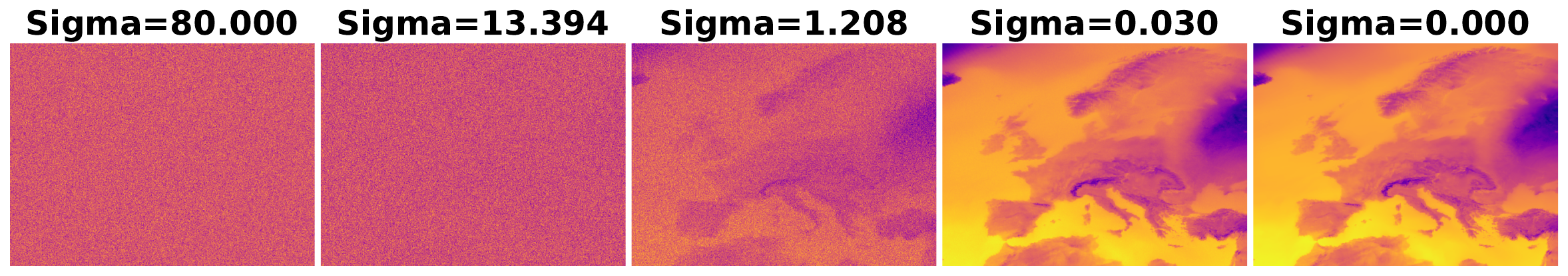}
        \par\small \textbf{(a)} EDM diffusion process.\\
        \label{fig:diffusion}
    \end{minipage}\hfill
    \begin{minipage}{\linewidth}
        \centering
        \includegraphics[width=\linewidth]{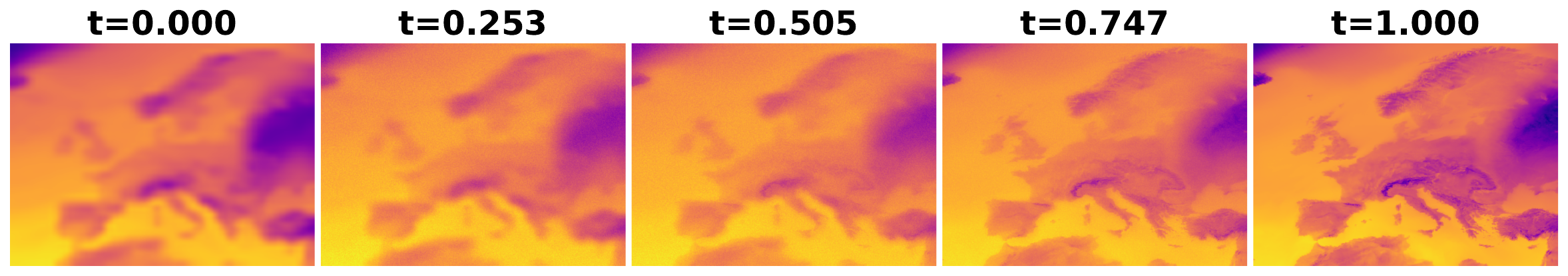}
        \par\small \textbf{(b)} Stochastic interpolant process.
        \label{fig:si}
    \end{minipage}
    \caption{Comparison of optimal trajectories for an EDM diffusion process and a stochastic interpolant between LQ and HQ. Unlike diffusion-based approaches that generate high-resolution fields from pure noise, our \acrshort{cdsi} constructs stochastic trajectories that evolve directly from the low-resolution input toward the target distribution, simplifying learning and improving sample realism.}
    \label{fig:Diffusion_vs_SI}
\end{figure*}

\subsection{Stochastic Interpolants}\label{sec:SI}
Stochastic interpolants were introduced by \citet{albergo2023stochastic}. The framework enables learning a stochastic differential equation (SDE) that implements a transport map between two distributions $p_0$ and $p_1$.
The framework has subsequently been extended to paired data $(x_0, x_1)$ \cite{SIfollmer}, which is the setting considered in this work. 

To design such an SDE, we first construct a stochastic interpolant
\begin{equation}
    x_t = \alpha(t) x_0 + \beta(t) x_1 + \sigma(t) W_t,    
\end{equation}
where $t \in [0,1]$ and $\alpha(t), \beta(t), \sigma(t) \in C^1(0,1)$ satify the boundary conditions $\alpha(0) = \beta(0) = 1$ and $\alpha(1) = \beta(0) = \sigma(1) = 0$. In our specific case we use $\alpha(t) = \sigma(t) = 1-t$, and $\beta(t) = t^2$. The data pair $(x_0, x_1)$ is sampled from the joint distribution $p(x_0, x_1)$ and $W_t$ is a Wiener process with $W \perp (x_0, x_1)$.

Let $b(t, x_t, x_0)$ denote the drift function that minimizes
\begin{align}
\mathcal{L}_b(\hat b)
=
\int_0^1
\mathbb{E}\!\left[
\left\lvert \hat b(t, x_t, x_0) - R_t \right\rvert^2
\right]
\, \diff t,\\
R_t
=
\dot{\alpha}(t)\, x_t
+
\dot{\beta}(t)\, x_0
+
\dot{\sigma}(t)\, W_t,
\end{align}
over all functions $\hat b(t, x_t, x_0)$, where $(x_0, x_1) \perp W_t$ and $W_t \stackrel{d}{=} \sqrt{t}z$ and $z \sim \mathcal{N}(0, I_d)$ at all $t \in [0,1]$.

Using this drift, we define the stochastic process $X_t$ by
\begin{align}
    \diff X_t = b(t, X_t, x_0) \diff t + \sigma(t) \diff W_t,\\
    X_{t=0}=x_0,
\end{align}

By construction, $\mathrm{Law}(X_t) = \mathrm{Law}(x_t|x_0)$ for all $(t, x_0) \in [0,1]$. The marginal distribution of $X_t$ matches the conditional distribution of $x_t$ given $x_0$ for all $t \in [0,1]$. Solving the SDE up to $t=1$ with different realizations of the Brownian motion $W$ produces samples $X_1$ from the target conditional distribution $p_{1 \mid x_0}$.

Before we can generate samples we need to learn the drift $b(t, X_t, x_0)$ which we do by approximating $b$ with a neural network $\hat{b}(t, X_t, x_0)$. The network is trained by minimizing the learning objective 
\begin{align}
    \mathcal{L} = \mathbb{E}_{x_0, x_1, t} \lVert \hat{b}(t, x_t, x_0) - R_t \rVert^2.
\end{align}

When we have approximated the drift $b$ with $\hat{b}$ we can generate new samples by numerically solving the SDE
\begin{align}
    \diff X_t = \hat{b}(t, X_t, x_0) \diff t + \sigma(t) \sqrt{t}z,\\
    z \sim \mathcal{N}(0, I_d), X_{t=0}=x_0.
\end{align}
by for example the Euler-Maruyama method.

\section{Method}
We follow the general framework of stochastic interpolants described in \cref{sec:SI} and tailor it to the task of downscaling global climate predictions to a high-resolution area of interest. We define $p_0$ as the distribution of LQ states and $p_{1 \mid x_0}$ as the distribution of HQ states conditioned on a specific LQ input, and construct a stochastic interpolant between these distributions. Intuitively, stochastic interpolants enable us to define generative trajectories that remain close to the data manifold throughout the sampling process. In contrast to diffusion models, which must learn to remove large amounts of noise before approaching the target distribution, stochastic interpolants start from a physically meaningful low-resolution state and introduce stochasticity only to model unresolved variability.

A comparison with trajectories obtained using EDM is shown in \cref{fig:Diffusion_vs_SI}. In contrast to EDM, the stochastic interpolant follows a trajectory that remains close to the data manifold, producing a stepwise transformation from the LQ input to the HQ output.

Since the input data from the \acrshort{esm} has a coarser resolution than the \acrshort{rcm} output, but our neural network architecture requires input and output to share the same resolution, we first upsample the \acrshort{esm} input using bilinear interpolation. This produces $x_0$, which matches the spatial resolution of the \acrshort{rcm} target. The target distribution $x_1$ is then given by the corresponding \acrshort{rcm} sample at the same time step. 

Since our goal is to perform conditional sampling of HQ fields given a LQ input, where the LQ input may contain additional fields beyond those being downscaled, we also include static features such as latitude/longitude coordinates, land–sea mask, and orography. These variables are provided to the drift model as a conditioning variable $C$ with the same spatial dimensions as $x_0$ and $x_1$ and are concatenated along the feature dimension.

To learn the drift $\hat{b}(t, x_t, x_0, C)$, we use a UNET implementation adapted from \citet{song2020score, karras2022elucidating}. Our UNET uses 128 feature channels at the top level and 256 channels at levels 2–4. The artificial time $t$ is encoded following \citet{karras2022elucidating}, using Fourier embeddings that transform $t$ into sine/cosine features at 128 frequencies with a base period of 16. These features are passed through a two-layer MLP with SiLU activation \citep{hendrycks2023gaussianerrorlinearunits}, producing a 512-dimensional representation. This time embedding is injected into the network via conditional layer norms in the MLP encoder and group norms in the UNET. The full model has approximately \SI{62.5}{M} parameters.

\begin{algorithm}[h!]
\caption{Sampling Algorithm}
\label{alg:sampling}
\begin{algorithmic}
\STATE \textbf{Input:} LQ sample $x_0 \sim p_0$, number of sampling steps $N$, conditioning variables $C$
\STATE $X_0 \gets x_0$
\STATE Define grid $0 = t_0 < t_1 < \dots < t_N = 1$
\STATE $\Delta t \gets t_1 - t_0$ (equal spacing)
\FOR{$n = 0$ to $N-1$}
    \STATE Sample $z \sim \mathcal{N}(0, I_d)$
    \STATE $X_{n+1} \gets X_n + \hat{b}(t_n, X_n, X_0, C)\,\Delta t + \sigma(t_n)\sqrt{\Delta t}\,z$
\ENDFOR
\STATE \textbf{Output:} $X_N$
\end{algorithmic}
\end{algorithm}

When sampling an ensemble member, we use a stochastic Euler-Maruyama sampling scheme as specified in \cref{alg:sampling} with 40 sampling steps. All ensemble members can be sampled in parallel, either with batched sampling on a single GPU or distributed over multiple GPUs, depending on the ensemble size.


\section{Experiments}

\begin{figure*}[t]      
  \centering
  \includegraphics[width=\textwidth,height=\textheight,keepaspectratio]{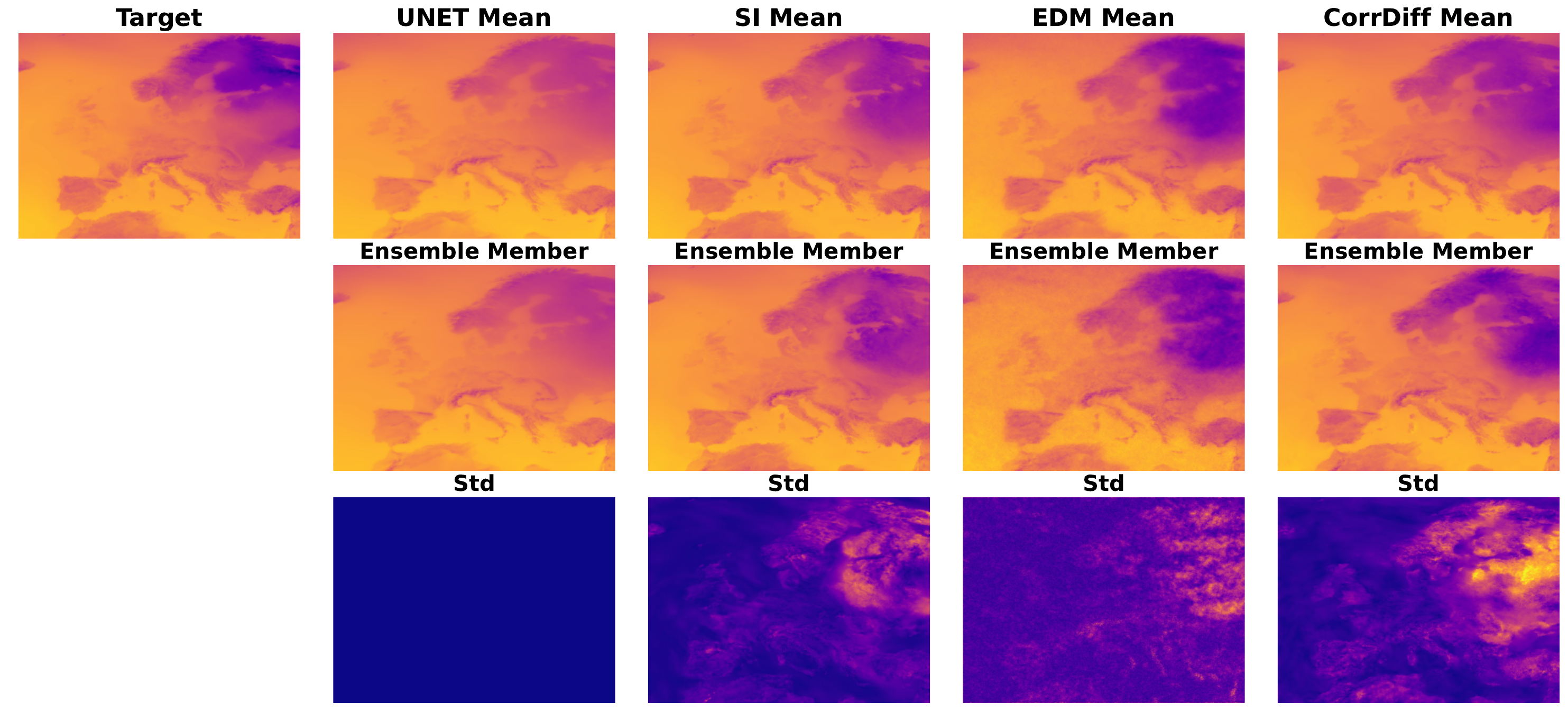}
  \caption{A qualitative comparison of the output for \SI{2}{\meter} temperature for all models. Note that the diffusion model is not able to remove all of the noise from the output despite using the same backbone architecture and training budget while the stochastic interpolant and CorrDiff models produces realistic samples. The UNET is a deterministic model and therefore the mean and member are the same and the same.}
  \label{fig:samples}
\end{figure*}

To assess model performance, we conduct experiments over both a validation period (see \cref{subsec:short_experiment}) and a test period (see \cref{subsec:long_experiment_future}) for all models. In addition, we examine the generalization performance of \acrshort{cdsi} using a previously unseen ESM–RCM realization in \cref{subsec:long_experiment_realization}. Additional evaluations are reported in \cref{apx:detailed_eval}.

\subsection{Metrics}
To evaluate our model, we use our model and the baselines to downscale \acrshort{esm} inputs and measure \acrfull{rmse}, \acrfull{ssr}, and \acrfull{crps}. In addition, we analyze the power spectra of the ensemble members to assess their physical realism. 
The detailed metric computations are explained in \cref{apx:metrics}. 

\subsection{Data}
We use the global \acrshort{esm} \acrshort{ec-earth}, available at a horizontal resolution of 1 degree (around \SI{110}{\kilo\meter}), as the coarse-scale input. The target model is the \acrshort{rcm} \acrshort{hclim}, which operates at a horizontal resolution of 0.1 degrees (around \SI{12}{\kilo\meter}).

The experiments are configured as follows. The model is trained on data from 1951--2008 using \acrshort{ec-earth} realization~1 as input and \acrshort{hclim} realization~1 as the target. Validation is performed on 2009 and the test period is 2010--2014 using the same realizations. An experiment covering 2010--2014 is conducted using \acrshort{ec-earth} realization~2 and \acrshort{hclim} realization~2, thereby testing generalization across both future climate scenarios and model realizations.

For all experiments, the model takes 24 input fields from the \acrshort{esm} and an additional 6 forcing features to predict precipitation and \SI{2}{\meter} air temperature as outputs. The input features are listed in \cref{tab:input_variables}.

Prior to training, all variables are normalized using the mean and standard deviation computed from the training period. Since the model backbone requires input and output fields to share the same spatial resolution, the \acrshort{ec-earth} inputs are upsampled to the \acrshort{hclim} grid using bilinear interpolation before being passed to the model.

\subsection{Baselines}
For our baselines we use the same UNET architecture for the backbone. We fix the training budget to 25 epochs. Training was performed on 1–4 A100 GPUs in a data-parallel setup with a learning rate of $10^{-5}$ and a batch size of 2. We use the AdamW optimizer \citep{adamw} with $\beta_1 = 0.9$, $\beta_2 = 0.95$, and weight decay $=0.1$. 

Our first baseline is a deterministic model trained with a mean squared error loss. Since it does not use an artificial diffusion time we remove the noise encoding and make the conditional normalization layers into standard normalization layers instead. Our second baseline is a conditional EDM diffusion model that directly models the distribution of the HQ conditioned on the LQ input. Our third baseline is CorrDiff, which combines a deterministic and a diffusion model in a two-stage process. In the first stage, a deterministic model predicts the conditional mean of the HQ field given the LQ input. In the second stage, an EDM diffusion model captures the residual uncertainty to reconstruct the full HQ distribution.

\paragraph{Computational Cost of SI vs.\ Baselines.}
To isolate differences in computational cost, we fix the backbone architecture across all models and focus on the number of function evaluations (NFEs) required to generate a HQ output. The deterministic UNET is the most computationally efficient model and requires only a single NFE to produce a prediction, but it can't produce ensemble forecasts.

For EDM and \acrshort{cdsi}, the number of NFEs is determined by the number of integration steps $N_{steps}$ used by the SDE or ODE solver. We generate \acrshort{cdsi} samples using an Euler–Maruyama scheme, requiring $N_{\text{steps}}$ NFEs per sample.  EDM employs a second-order solver with 20 integration steps, resulting in $N_{steps} \times 2 - 1$ NFEs per sample. CorrDiff uses the same diffusion solver as EDM but additionally requires a deterministic UNET forward pass to predict the conditional mean, yielding a total of $N_{steps} \times 2$ NFEs per sample.

While matching the number of NFEs allows the probabilistic models to achieve comparable inference times, their training and memory costs could differ substantially. CorrDiff requires training two separate models, a deterministic UNET for mean prediction and a diffusion model for residuals, potentially increasing the overall training cost compared to single-model approaches. This can become particularly burdensome when working with large datasets or models.

In addition, CorrDiff has a substantially higher memory footprint, since both the deterministic UNET and the diffusion model must reside in memory during training and inference. One workaround is to precompute UNET predictions for the training set, but this doubles storage and adds complexity. At inference, the sequential execution of both models also requires significant memory, potentially limiting model, batch, or data sizes. Strategies such as distributing models across GPUs or swapping between CPU and GPU can reduce memory use, but at the cost of slower inference and more complex deployment.


\subsection{Validation period: 2009 (\acrshort{ec-earth} realization 1, \acrshort{hclim} realization 1)}
\label{subsec:short_experiment}
We use the year 2009 to compare the baseline methods and to conduct ablation studies of our model. This short evaluation period is chosen due to the high computational cost associated with evaluating multiple models over longer time spans. We evaluate the performance of CorrDiff and \acrshort{cdsi} across ensemble sizes and NFEs. As EDM fails to generate realistic ensemble members, we omit it from evaluations with larger ensemble sizes and from the longer experiments.

Overall, we observe that the ensemble mean RMSE generally improves for precipitation as the number of ensemble members increases. For temperature, however, performance plateaus once the ensemble size exceeds 20. The results are summarized in \cref{tab:model_metrics_2009}. 

\begin{table}[h!]
  \centering
  \caption{Verification metrics for the test period 2009.}
  \label{tab:model_metrics_2009}
  \sisetup{
    detect-weight=true,
    detect-family=true,
    table-format=3.3
  }
  \setlength{\tabcolsep}{3pt} 
  \begin{tabular*}{\linewidth}{@{\extracolsep{\fill}} l S S S @{}}
    \toprule
    Model & {RMSE} & {SSR} & {CRPS} \\
    \midrule
    
    \multicolumn{4}{l}{\textbf{Precipitation}} \\
    \addlinespace[0.3em]
    UNET & 3.183 & {--} & {--} \\
    EDM 39 NFE 5 ens & 3.992 & 0.952 & 1.402\\
    CorrDiff 40 NFE 5 ens & 3.494 & 0.909 & 1.146\\
    CorrDiff 40 NFE 20 ens & 3.319 & 0.894 & 1.146\\
    CorrDiff 40 NFE 50 ens & 3.272 & 0.891 & 1.146\\
    CorrDiff 40 NFE 100 ens & 3.259 & 0.892 & 1.146\\
    SI 40 NFE 5 ens & 3.396 & 0.715 & 1.159\\
    SI 40 NFE 20 ens & 3.281 & 0.690 & 1.159\\
    SI 40 NFE 50 ens & 3.255 & 0.685 & 1.159\\
    SI 40 NFE 100 ens & 3.246 & 0.684 & 1.155\\
        
    \addlinespace[0.6em]
    
    \multicolumn{4}{l}{\textbf{Temperature (2\,m)}}\\
    \addlinespace[0.3em]
    UNET & 1.870 & {--} & {--} \\
    EDM 39 NFE 5 ens & 2.943 & 0.724 & 1.453\\
    CorrDiff 40 NFE 5 ens & 2.166 & 0.638 & 1.000\\
    CorrDiff 40 NFE 20 ens & 2.110 & 0.614 & 1.000\\
    CorrDiff 40 NFE 50 ens & 2.101 & 0.607 & 1.000\\
    CorrDiff 40 NFE 100 ens & 2.101 & 0.605 & 1.000\\
    SI 40 NFE 5 ens & 2.036 & 0.629 & 0.898\\
    SI 40 NFE 20 ens & 1.971 & 0.606 & 0.888\\
    SI 40 NFE 50 ens & 1.981 & 0.597 & 0.898\\
    SI 40 NFE 100 ens & 1.971 & 0.597 & 0.898\\  
    \bottomrule
  \end{tabular*}
\end{table}

In \cref{tab:model_metrics_2009_NFE_ablation}, we report an ablation study on the NFEs. As expected, all models generally improve in terms of RMSE and CRPS as the NFEs increases , with diminishing returns beyond a around 40 NFEs. For \acrshort{cdsi}, increasing the number of solver steps has a pronounced effect on ensemble spread and calibration. When using too few steps (e.g., 10 NFEs), the ensemble is clearly underdispersed, leading to low SSR values. As the number of steps increases, the ensemble spread grows and the SSR improves substantially, indicating better calibration. In contrast, CorrDiff exhibits the opposite trend. While RMSE continues to decrease slightly with more NFEs, the SSR systematically deteriorates as the number of steps increases, suggesting that the ensemble becomes increasingly underdispersed.

While the deterministic UNET performs well in terms of RMSE, it fails to reproduce realistic spectral properties and does not support ensemble generation for uncertainty quantification, motivating the use of generative models for climate downscaling.

\begin{figure}[h!]
    \centering
    \begin{minipage}{\linewidth}
        \centering
        \includegraphics[width=\linewidth]{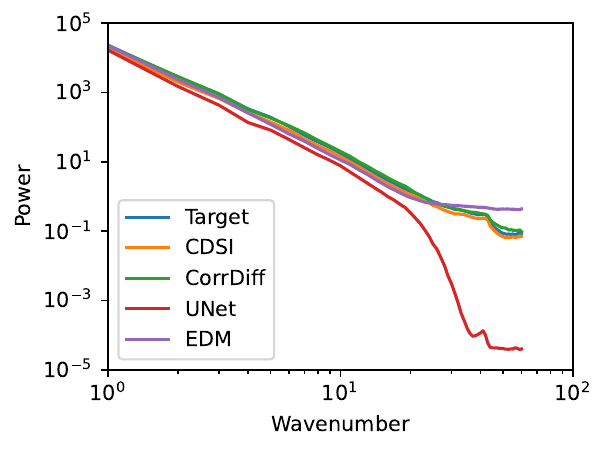}
        \par\small \textbf{(a)} Precipitation spectrum \\
        \label{fig:precip_spectrum}
    \end{minipage}\hfill
    \begin{minipage}{\linewidth}
        \centering
        \includegraphics[width=\linewidth]{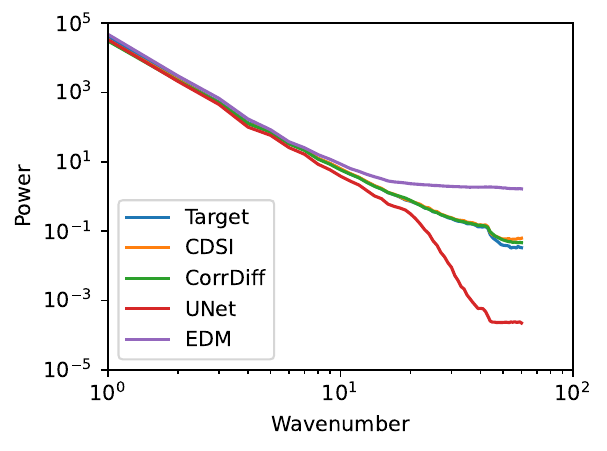}
        \par\small \textbf{(b)} Temperature spectrum \\
        \label{fig:temp_spectrum}
    \end{minipage}
    \caption{Power spectra of temperature and precipitation.}
    \label{fig:spectra}
\end{figure}

Examining the ensemble members and ensemble standard deviation for EDM in \cref{fig:samples} and the power spectra in \cref{fig:spectra}, we observe that, under the same training and parameter budget, the diffusion model struggles to fully remove high-frequency noise from its outputs. We argue that the improved performance of the stochastic interpolant formulation stems from the learned trajectory evolving from LQ inputs toward HQ targets, rather than from pure noise to HQ targets as in diffusion-based approaches. This change in trajectory may simplify the learning problem, since the LQ input provides a substantially more informative prior than Gaussian noise. Moreover, if the generation process is terminated slightly early, the resulting sample lies between LQ and HQ, rather than between noise and HQ. In practice, such intermediate samples tend to appear as mildly smoothed versions of the target and are therefore visually and physically more realistic than samples that retain residual noise. This distinction is shown in \cref{fig:Diffusion_vs_SI}.


\subsection{Test period: 2010--2014 (\acrshort{ec-earth} realization 1, \acrshort{hclim} realization 1)}
\label{subsec:long_experiment_future}
This experiment evaluates the model’s ability to generalize to future climate conditions that are not observed during training, while keeping the underlying climate model realizations fixed. For the longer experiments, we restrict the number of function evaluations to 40 and use an ensemble size of 20 for due to computational constraints. 

As shown in \cref{tab:model_metrics_2009_2014}, \acrshort{cdsi} generally attains slightly lower ensemble-mean RMSE than CorrDiff. For the spread–skill ratio, the relative performance depends on the variable, where CorrDiff is better calibrated for precipitation and \acrshort{cdsi} shows improved calibration for temperature. In terms of the continuous ranked probability score, the two methods exhibit comparable performance for precipitation, while \acrshort{cdsi} performs better for temperature.

\begin{table}[h!]
  \centering
  \caption{Verification metrics for the test period 2010--2014 for realization 1.}
  \label{tab:model_metrics_2009_2014}
  \sisetup{
    detect-weight=true,
    detect-family=true,
    table-format=3.3
  }
  \begin{tabular*}{\linewidth}{@{\extracolsep{\fill}} l S S S @{}}
    \toprule
    Model & {RMSE} & {SSR} & {CRPS} \\
    \midrule

    \multicolumn{4}{l}{\textbf{Precipitation}} \\
    \addlinespace[0.3em]
    UNET     & 3.070 & {--} & {--} \\
    CorrDiff & 3.209 & 0.896 & 1.115 \\
    SI       & 3.171 & 0.693 & 1.118\\
    \addlinespace[0.6em]

    \multicolumn{4}{l}{\textbf{Temperature (2\,m)}} \\
    \addlinespace[0.3em]
    UNET     & 1.603 & {--} & {--} \\
    CorrDiff & 1.837 & 0.702 & 0.925 \\
    SI       & 1.631 & 0.745 & 0.788 \\
    \bottomrule
  \end{tabular*}
\end{table}




\subsection{Test period: 2010--2014 (\acrshort{ec-earth} realization 2, \acrshort{hclim} realization 2)}
\label{subsec:long_experiment_realization}
In this experiment, we assess the sensitivity of the model to the choice of climate model realization. Specifically, we investigate whether a model trained on one realization can generalize to a different realization and thereby capture climate variability beyond the conditions seen during training. As in \cref{subsec:long_experiment_future} we restrict the number of function evaluations to 40 and use an ensemble size of 20. 

As shown in \cref{tab:model_metrics_2009_2014_r2}, the models exhibit broadly comparable performance across most evaluation metrics, with \acrshort{cdsi} showing a clear improvement in temperature RMSE, even outperforming the UNET. Overall, the results indicate that all considered models generalize well to a new \acrshort{esm}–\acrshort{rcm} realization, with no substantial degradation in performance.

\begin{table}[h!]
  \centering
  \caption{Verification metrics for the test period 2010--2014 for realization 2.}
  \label{tab:model_metrics_2009_2014_r2}
  \sisetup{
    detect-weight=true,
    detect-family=true,
    table-format=3.3
  }
  \begin{tabular*}{\linewidth}{@{\extracolsep{\fill}} l S S S @{}}
    \toprule
    Model & {RMSE} & {SSR} & {CRPS} \\
    \midrule

    \multicolumn{4}{l}{\textbf{Precipitation}} \\
    \addlinespace[0.3em]
    UNET     & 3.070 & {--} & {--} \\
    CorrDiff & 3.226 & 0.893 & 1.112 \\
    SI       & 3.199 & 0.695 & 1.120 \\
    \addlinespace[0.6em]

    \multicolumn{4}{l}{\textbf{Temperature (2\,m)}} \\
    \addlinespace[0.3em]
    UNET     & 1.603 & {--} & {--} \\
    CorrDiff & 1.742 & 0.726 & 0.890 \\
    SI       & 1.533 & 0.766 & 0.744 \\
    \bottomrule
  \end{tabular*}
\end{table}

\section{Conclusion}
\label{sec:conclusion}

In this work, we introduced \acrfull{cdsi} and demonstrated its ability to emulate high-resolution regional climate simulations from coarse-resolution \acrshort{esm} at a fraction of the computational cost of a \acrshort{rcm}. We further demonstrated that \acrshort{cdsi} generalizes beyond the conditions seen during training. The model maintains high skill when evaluated on unseen future periods and on new climate model realizations, indicating robustness to both temporal distribution shift and realization uncertainty. While \acrshort{cdsi} produces samples that are visually and spectrally realistic, the overall verification metrics are similar to those of CorrDiff. Both methods outperform standard diffusion models. A practical advantage of \acrshort{cdsi} is that it attains comparable performance to CorrDiff using a single probabilistic model, without relying on intermediate \acrshort{rcm} downscaling or additional deterministic components.

Overall, these results illustrate the potential of stochastic interpolants for probabilistic climate downscaling, enabling large ensembles and long-horizon simulations that are currently very computationally expensive with \acrshort{rcm}s. Given the same network backbone and training budget, \acrshort{cdsi} can produce realistic samples with a simplified workflow, making it a viable alternative to multi-stage diffusion-based approaches.

Future work could focus on extending the framework to downscale more variables, incorporating explicit physical constraints, improving the representation of extremes, and evaluating performance across new driving \acrshort{esm}s and emission scenarios. 



\section*{Acknowledgements}
This research is financially supported by the Swedish Research Council (grant no: 2020-04122, 2024-05011), 
the Wallenberg AI, Autonomous Systems and Software Program (WASP) funded by the Knut and Alice Wallenberg Foundation,
and the Excellence Center at Linköping--Lund in Information Technology (ELLIIT).
Our computations were enabled by the Berzelius resource at the National Supercomputer Centre, provided by the  Knut and Alice Wallenberg Foundation.
This paper was partially funded by the AI4PEX project (AI4PEX – Artificial Intelligence and Machine Learning for Enhanced Representation of Processes and Extremes in Earth System Models), which has received funding from the European Union's Horizon Europe research and innovation programme under grant agreement no. 101137682; the UK Research and Innovation (UKRI) under the UK government's Horizon Europe funding guarantee under grant agreement numbers 10114295, 10103109, and 10093450, and the Swiss State Secretariat for Education, Research and Innovation (SERI) under grant numbers 23.00546 and 24.00178.

\section*{Impact Statement}
This work contributes to the development of data-driven methods for climate downscaling, with the goal of improving access to high-resolution regional climate information. By substantially reducing the computational cost of generating regional climate ensembles compared to RCMs, the proposed approach has the potential to support larger ensemble studies and a broader exploration of climate scenarios. These capabilities are relevant for climate impact assessments, risk analysis, and adaptation planning, where high-resolution information is often required, but computationally constrained.




\bibliography{biblio}
\bibliographystyle{icml2026}

\newpage
\appendix
\onecolumn



\section{Data}
Here we provide additional details about the input data to the model with all inputs listed in \cref{tab:input_variables}.
\begin{table}[ht]
\centering
\caption{Input variables and units}
\label{tab:input_variables}
\begin{tabular}{lll}
\hline
Variable & Description & Unit \\
\hline
& \textbf{ESM Output} & \\
cloud\_cover & Cloud fraction & [0,1] \\
humidity\_1000 & Spec. humidity at 1000 hPa & kg kg$^{-1}$ \\
humidity\_850 & Spec. humidity at 850 hPa & kg kg$^{-1}$ \\
humidity\_700 & Spec. humidity at 700 hPa & kg kg$^{-1}$ \\
humidity\_500 & Spec. humidity at 500 hPa & kg kg$^{-1}$ \\
humidity\_250 & Spec. humidity at 250 hPa & kg kg$^{-1}$ \\
precipitation & Accumulated precipitation & mm \\
sea\_level\_pressure & Sea-level pressure & hPa \\
temperature\_2m & 2 m air temperature & K \\
temperature\_1000 & Temperature at 1000 hPa & K \\
temperature\_850 & Temperature at 850 hPa & K \\
temperature\_700 & Temperature at 700 hPa & K \\
temperature\_500 & Temperature at 500 hPa & K \\
temperature\_250 & Temperature at 250 hPa & K \\
wind\_u\_1000 & Zonal wind at 1000 hPa & m s$^{-1}$ \\
wind\_u\_850 & Zonal wind at 850 hPa & m s$^{-1}$ \\
wind\_u\_700 & Zonal wind at 700 hPa & m s$^{-1}$ \\
wind\_u\_500 & Zonal wind at 500 hPa & m s$^{-1}$ \\
wind\_u\_250 & Zonal wind at 250 hPa & m s$^{-1}$ \\
wind\_v\_1000 & Meridional wind at 1000 hPa & m s$^{-1}$ \\
wind\_v\_850 & Meridional wind at 850 hPa & m s$^{-1}$ \\
wind\_v\_700 & Meridional wind at 700 hPa & m s$^{-1}$ \\
wind\_v\_500 & Meridional wind at 500 hPa & m s$^{-1}$ \\
wind\_v\_250 & Meridional wind at 250 hPa & m s$^{-1}$ \\
\hline
& \textbf{Static and forcing features} & \\
oro & Orography & m \\
lsm & Land-sea-mask & [0,1] \\
x\_coord & x-coordinate & [0,1] \\
y\_coord & y-coordinate & [0,1] \\
toy\_cos & Time of year cosine & [-1,1] \\
toy\_sine & Time of year sine & [-1,1] \\
\hline
\end{tabular}
\end{table}

\section{Metrics}\label{apx:metrics}
Given an ensemble forecast $\hat{X}$, forecast accuracy is quantified using the root mean squared error (RMSE). For variable $d$ at lead time $t$, the RMSE is computed from the ensemble-mean prediction $\bar{\hat{X}}^{t}_{g,d}$ over all spatial grid points $g \in G$ as
\[
\text{RMSE}^t_d
= \left( \frac{1}{|G|} \sum_{g \in G}
\left( \bar{\hat{X}}^{t}_{g,d} - X^{t}_{g,d} \right)^2
\right)^{1/2}.
\]
The ensemble mean prediction is defined by
\[
\bar{\hat{X}}^{t}_{g,d}
= \frac{1}{N_{\text{ens}}}
\sum_{\text{ens}=1}^{N_{\text{ens}}}
\hat{X}^{t}_{g,d,\text{ens}},
\]
where $\hat{X}^{t}_{g,d,\text{ens}}$ denotes the forecast from ensemble member $\text{ens}$ and $N_{\text{ens}}$ is the total number of ensemble members. Following standard practice and the WeatherBench~2 benchmark \citep{rasp2023weatherbench}, the spatial averaging is performed prior to taking the square root.

To evaluate the calibration of the ensemble uncertainty, we employ the bias-corrected spread--skill ratio (SSR). For variable $d$ at time $t$, the SSR is given by
\[
\text{SSR}^t_d
= \sqrt{\frac{N_{\text{ens}}+1}{N_{\text{ens}}}}
\frac{\text{Spread}^t_d}{\text{RMSE}^t_d}.
\]
The ensemble spread is computed as
\[
\text{Spread}^t_d
= \left(
\frac{1}{|G| N_{\text{ens}}}
\sum_{g \in G}
\sum_{\text{ens}=1}^{N_{\text{ens}}}
\left(
\hat{X}^{t}_{g,d,\text{ens}} - \bar{\hat{X}}^{t}_{g,d}
\right)^2
\right)^{1/2}.
\]
Values of $\text{SSR}^t_d$ close to one indicate well-calibrated predictive uncertainty \citep{fortin2014should}.

In addition, probabilistic forecast performance is assessed using the continuous ranked probability score (CRPS) \citep{gneiting2007strictly}. For variable $d$ at lead time $t$, the CRPS is estimated as
\[
\text{CRPS}^t_d
= \frac{1}{|G| N_{\text{ens}}}
\sum_{g \in G}
\left(
\sum_{\text{ens}=1}^{N_{\text{ens}}}
\left|
\hat{X}^{t}_{g,d,\text{ens}} - X^{t}_{g,d}
\right|
-
\frac{1}{2(N_{\text{ens}}-1)}
\sum_{\text{ens}=1}^{N_{\text{ens}}}
\sum_{\text{ens}^*=1}^{N_{\text{ens}}}
\left|
\hat{X}^{t}_{g,d,\text{ens}}
-
\hat{X}^{t}_{g,d,\text{ens}^*}
\right|
\right).
\]
This expression corresponds to a finite-sample estimator of the CRPS \citep{zamo2018estimation} and treats ensemble members independently, without explicitly accounting for their covariance structure.

When computing verification metrics for individual variables, all predictions are first transformed back to their original physical units before being compared with the ground truth.

The Mathews Correlation Coefficient formula is given by:
\begin{equation}
\mathrm{MCC} = 
\frac{TP \times TN - FP \times FN}
{\sqrt{(TP + FP)(TP + FN)(TN + FP)(TN + FN)}}
\end{equation}
 
where TP and TN denote the number of true positives and true negatives (correctly identified exceedance and non-exceedance days), and FP and FN represent false positives and false negatives, respectively. The coefficient ranges from $-1$ (complete disagreement) to $+1$ (perfect agreement), with 0 indicating random correspondence.

\section{Detailed Evaluation}\label{apx:detailed_eval}
In this section, we present additional evaluation results that provide a more detailed characterization of model performance. We first study the impact of the number of function evaluations on validation performance in \cref{tab:model_metrics_2009_NFE_ablation}. We then examine the temporal evolution of the verification metrics to assess their stability over the evaluation period. Next, we analyze how model performance varies with the magnitude of the target variables, and finally relate the probabilistic metrics to the error of the deterministic UNET baseline.

\begin{table}[h!]
  \centering
  \caption{Ablation of the number of function evaluations, evaluated on verification metrics for the test period 2009 with 5 ensemble members.}
  \label{tab:model_metrics_2009_NFE_ablation}
  \sisetup{
    detect-weight=true,
    detect-family=true,
    table-format=3.3
  }
  \begin{tabular*}{\linewidth}{@{\extracolsep{\fill}} l S S S @{}}
    \toprule
    Model & {RMSE} & {SSR} & {CRPS} \\
    \midrule
    
    \multicolumn{4}{l}{\textbf{Precipitation}} \\
    \addlinespace[0.3em]
    EDM 9 NFE & 5.174 & 1.415 & 1.964 \\
    EDM 19 NFE & 4.031 & 0.980 & 1.415 \\
    EDM 39 NFE & 3.992 & 0.952 & 1.402\\
    EDM 99 NFE & 3.982 & 0.930 & 1.399 \\
    CorrDiff 10 NFE & 3.834 & 1.269 & 1.306 \\
    CorrDiff 20 NFE & 3.566 & 0.985 & 1.161 \\
    CorrDiff 40 NFE & 3.494 & 0.909 & 1.146\\
    CorrDiff 100 NFE & 3.492 & 0.884 & 1.149 \\
    SI 10 NFE & 3.787 & 0.430 & 1.357 \\
    SI 20 NFE & 3.442 & 0.579 & 1.197 \\
    SI 40 NFE & 3.396 & 0.715 & 1.159\\
    SI 100 NFE & 3.481 & 0.820 & 1.167 \\
        
    \addlinespace[0.6em]
    
    \multicolumn{4}{l}{\textbf{Temperature (2\,m)}}\\
    \addlinespace[0.3em]
    EDM 9 NFE & 6.272 & 1.397 & 3.054 \\
    EDM 19 NFE & 3.135 & 0.808 & 1.556 \\
    EDM 39 NFE & 2.943 & 0.724 & 1.453\\
    EDM 99 NFE & 2.928 & 0.703 & 1.446 \\
    CorrDiff 10 NFE & 2.983 & 1.117 & 1.371 \\
    CorrDiff 20 NFE & 2.239 & 0.725 & 1.013 \\
    CorrDiff 40 NFE & 2.166 & 0.638 & 1.000\\
    CorrDiff 100 NFE & 2.150 & 0.624 & 0.995 \\
    SI 10 NFE & 3.481 & 0.393 & 1.766 \\
    SI 20 NFE & 2.282 & 0.529 & 1.020 \\
    SI 40 NFE & 2.036 & 0.629 & 0.898 \\
    SI 100 NFE & 2.041 & 0.678 & 0.922 \\
    \bottomrule
  \end{tabular*}
\end{table}

\begin{figure}[h!]
    \centering
    \includegraphics[width=\linewidth]{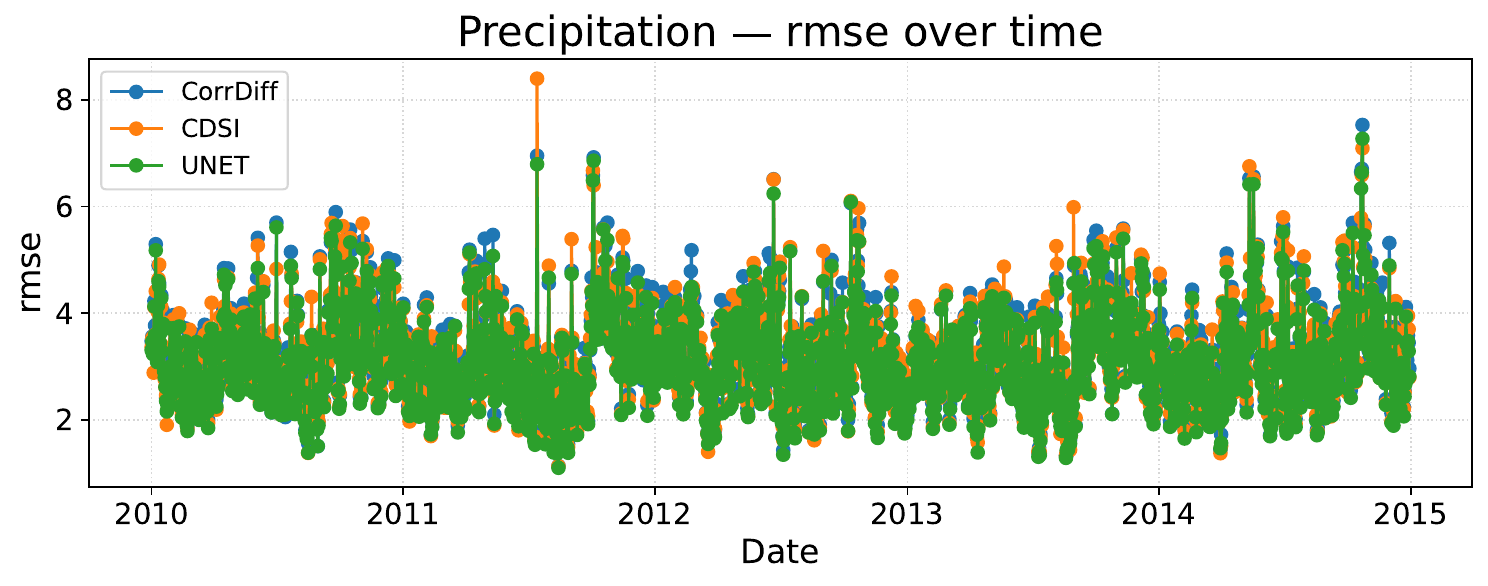}
    \caption{The RMSE over time for precipitation for the test period 2010--2014 for realization 1}
    \label{fig:rmse_time_pr}
\end{figure}

\begin{figure}[h!]
    \centering
    \includegraphics[width=\linewidth]{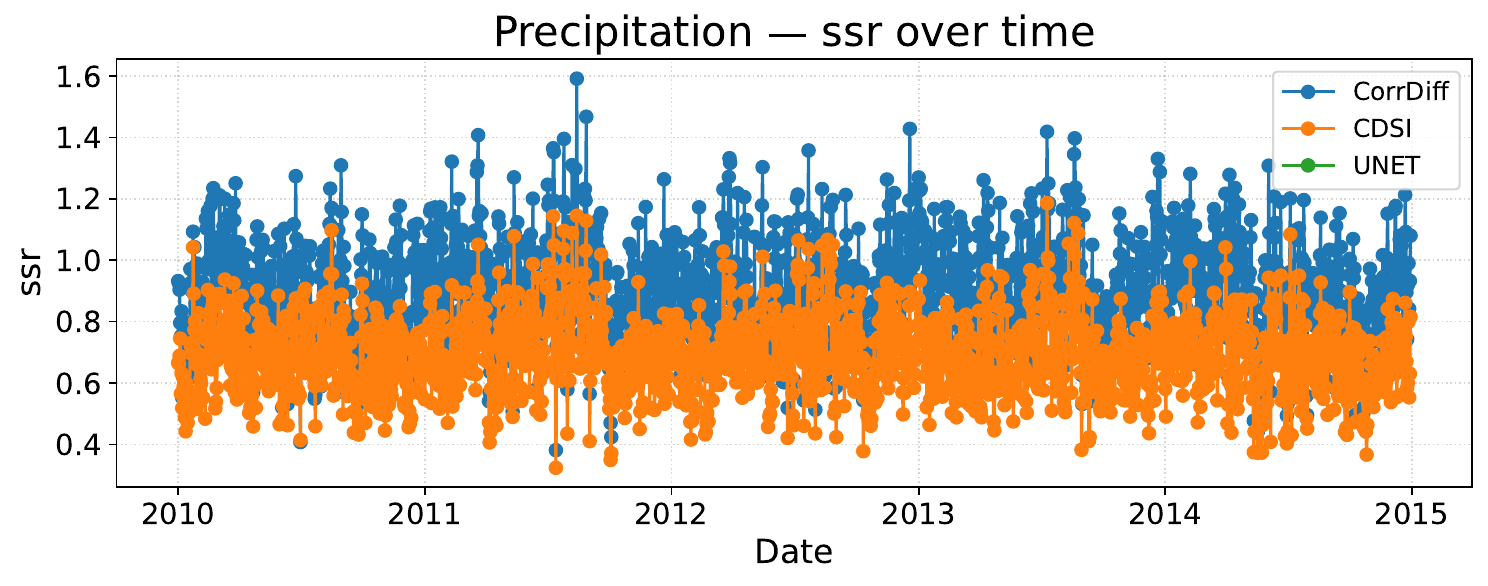}
    \caption{The SSR over time for precipitation for the test period 2010--2014 for realization 1}
    \label{fig:ssr_time_pr}
\end{figure}

\begin{figure}[h!]
    \centering
    \includegraphics[width=\linewidth]{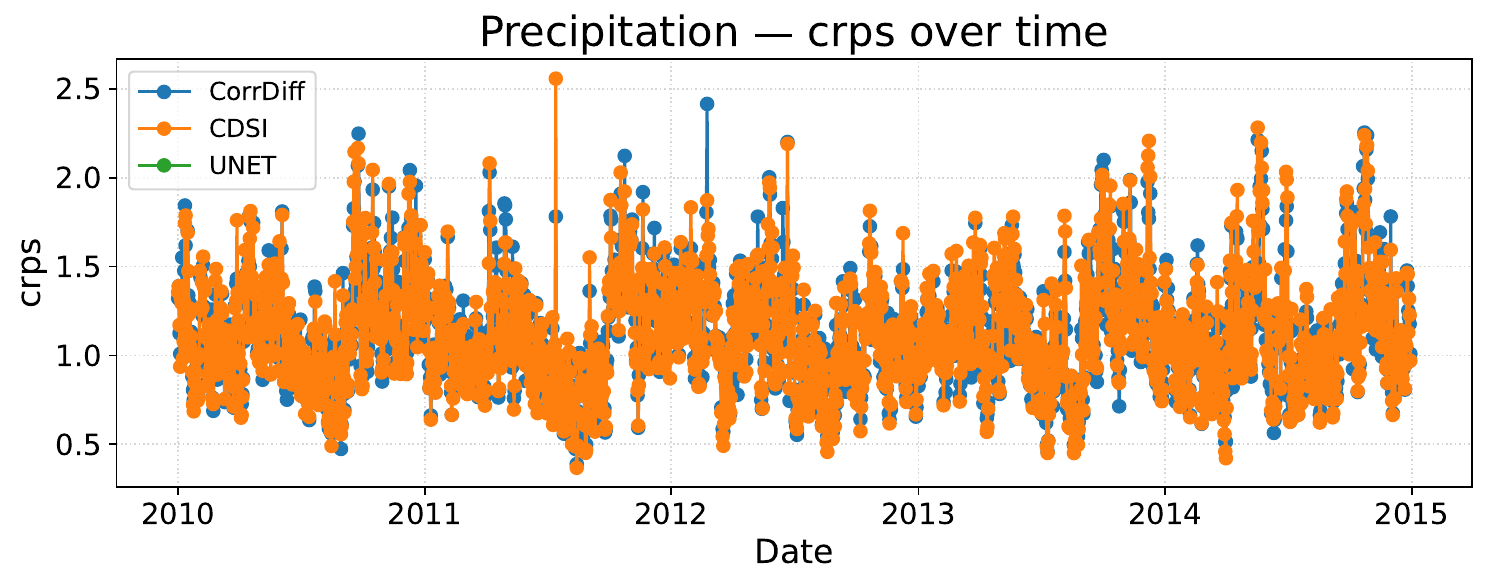}
    \caption{The CRPS over time for precipitation for the test period 2010--2014 for realization 1}
    \label{fig:crps_time_pr}
\end{figure}

\begin{figure}[h!]
    \centering
    \includegraphics[width=\linewidth]{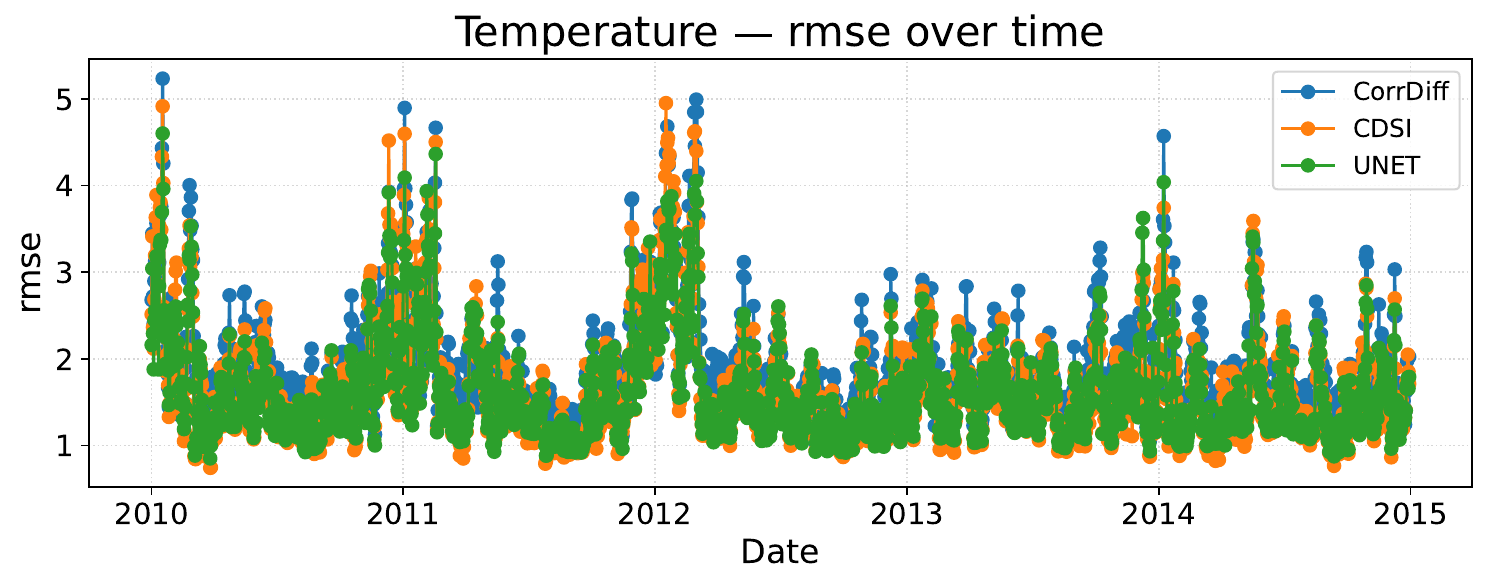}
    \caption{The RMSE over time for temperature for the test period 2010--2014 for realization 1}
    \label{fig:rmse_time_tas}
\end{figure}

\begin{figure}[h!]
    \centering
    \includegraphics[width=\linewidth]{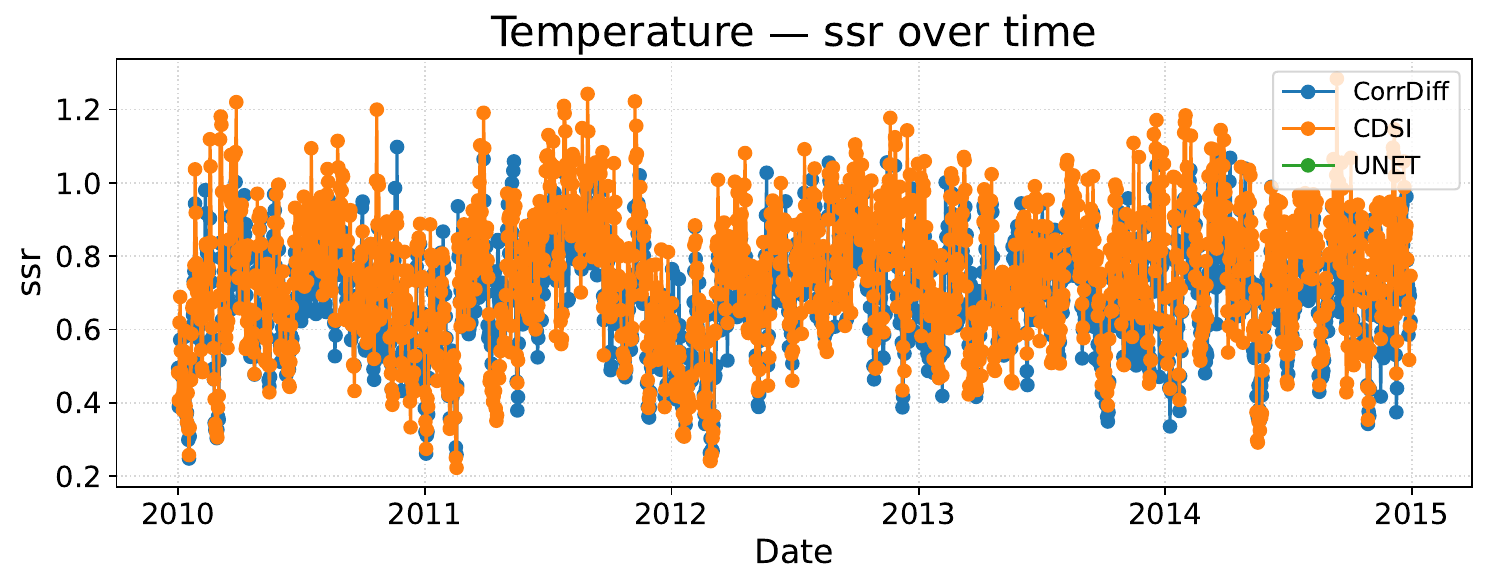}
    \caption{The SSR over time for temperature for the test period 2010--2014 for realization 1}
    \label{fig:ssr_time_tas}
\end{figure}

\begin{figure}[h!]
    \centering
    \includegraphics[width=\linewidth]{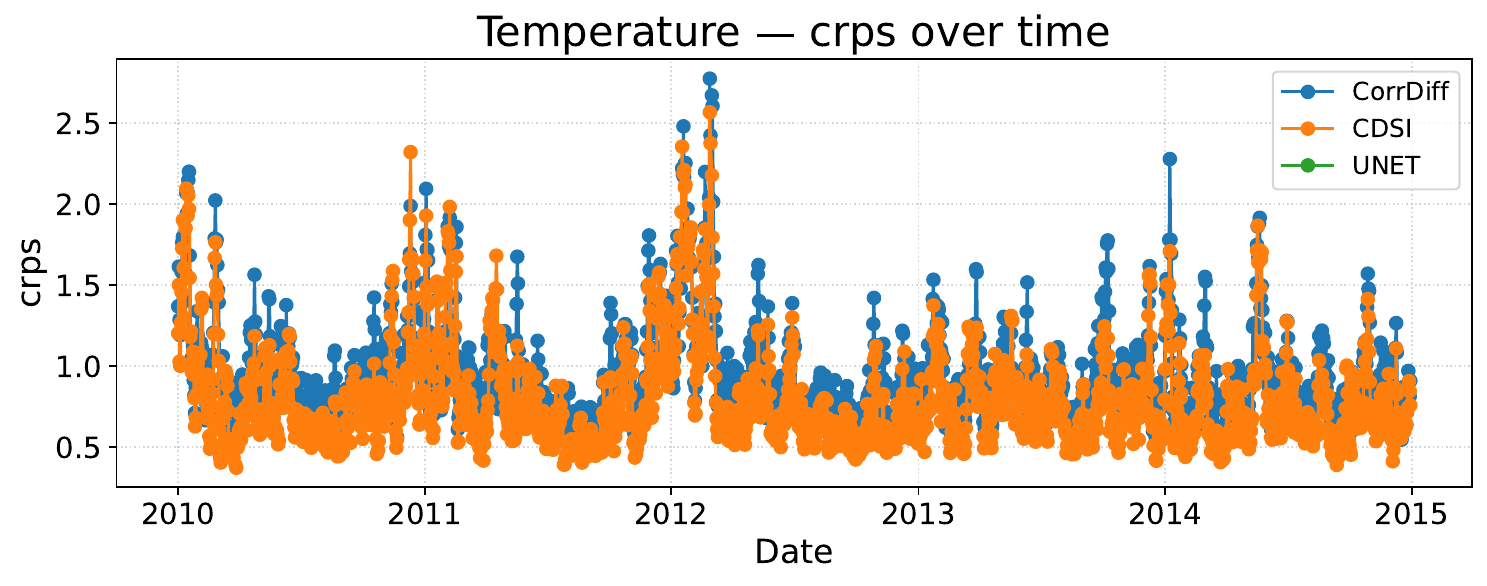}
    \caption{The CRPS over time for temperature for the test period 2010--2014 for realization 1}
    \label{fig:crps_time_tas}
\end{figure}

Figures \cref{fig:rmse_time_pr,fig:ssr_time_pr,fig:crps_time_pr} show the time series of RMSE, SSR, and CRPS for precipitation over the test period, while \cref{fig:rmse_time_tas,fig:ssr_time_tas,fig:crps_time_tas} present the corresponding results for temperature. These plots highlight the temporal variability of model skill and calibration.

\begin{figure}[h!]
    \centering
    \includegraphics[width=0.32\linewidth]{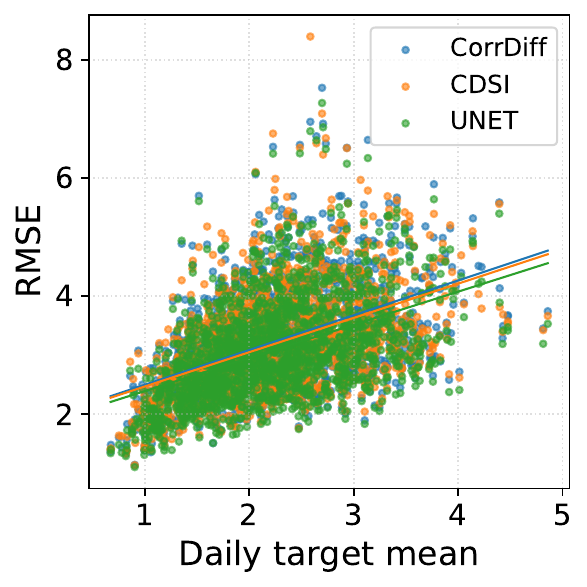}
    \includegraphics[width=0.32\linewidth]{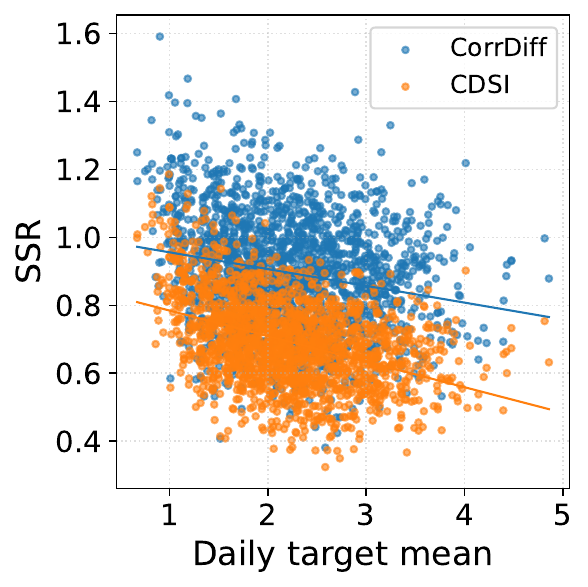}
    \includegraphics[width=0.32\linewidth]{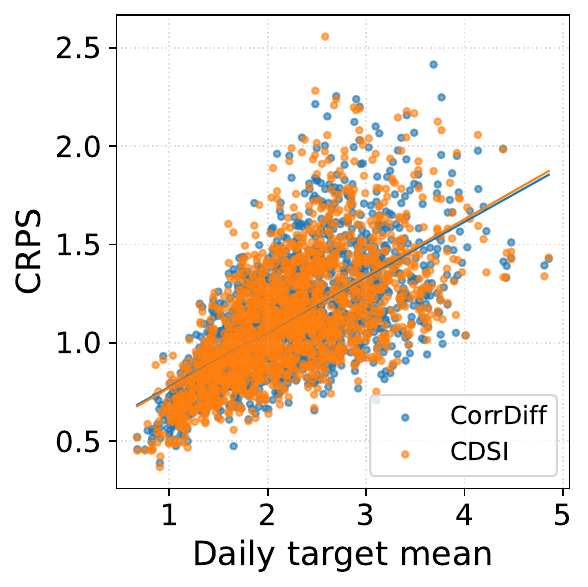}
    \caption{The RMSE, SSR, and CRPS as a function of the mean precipitation for the test period 2010--2014 for realization 1}
    \label{fig:metrics_vs_truth_pr}
\end{figure}

\begin{figure}[h!]
    \centering
    \includegraphics[width=0.32\linewidth]{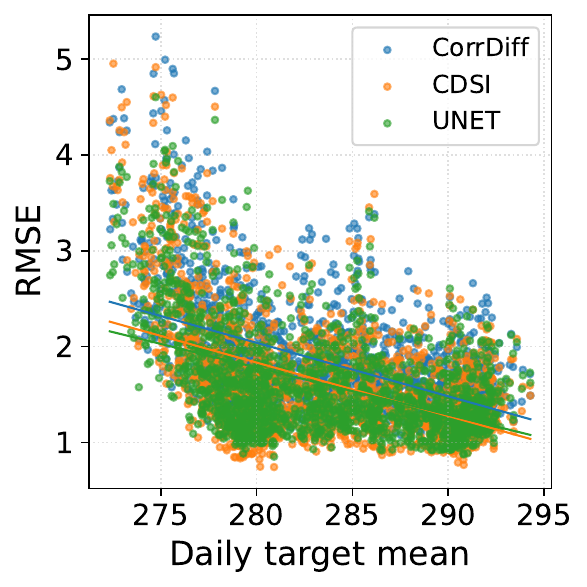}
    \includegraphics[width=0.32\linewidth]{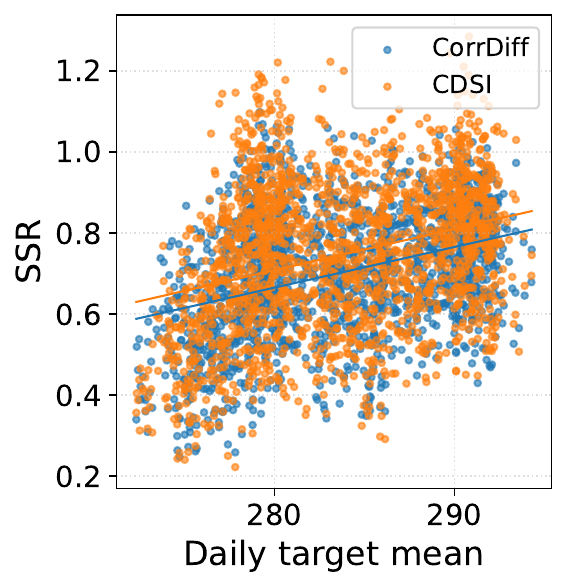}
    \includegraphics[width=0.32\linewidth]{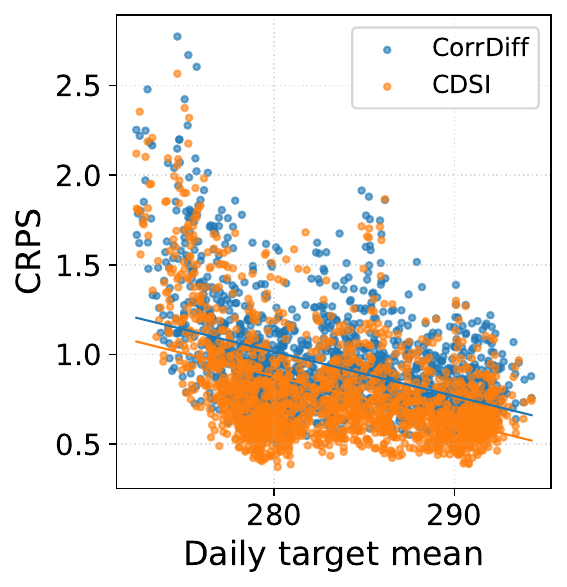}
    \caption{The RMSE, SSR, and CRPS as a function of the mean temperature for the test period 2010--2014 for realization 1}
    \label{fig:metrics_vs_truth_tas}
\end{figure}

To investigate how performance varies with the underlying climate state, \cref{fig:metrics_vs_truth_pr,fig:metrics_vs_truth_tas} report RMSE, SSR, and CRPS as functions of the mean precipitation and temperature, respectively. This analysis helps assess whether errors or miscalibration are associated with particular regimes.

\begin{figure}[h!]
    \centering
    \includegraphics[width=\linewidth]{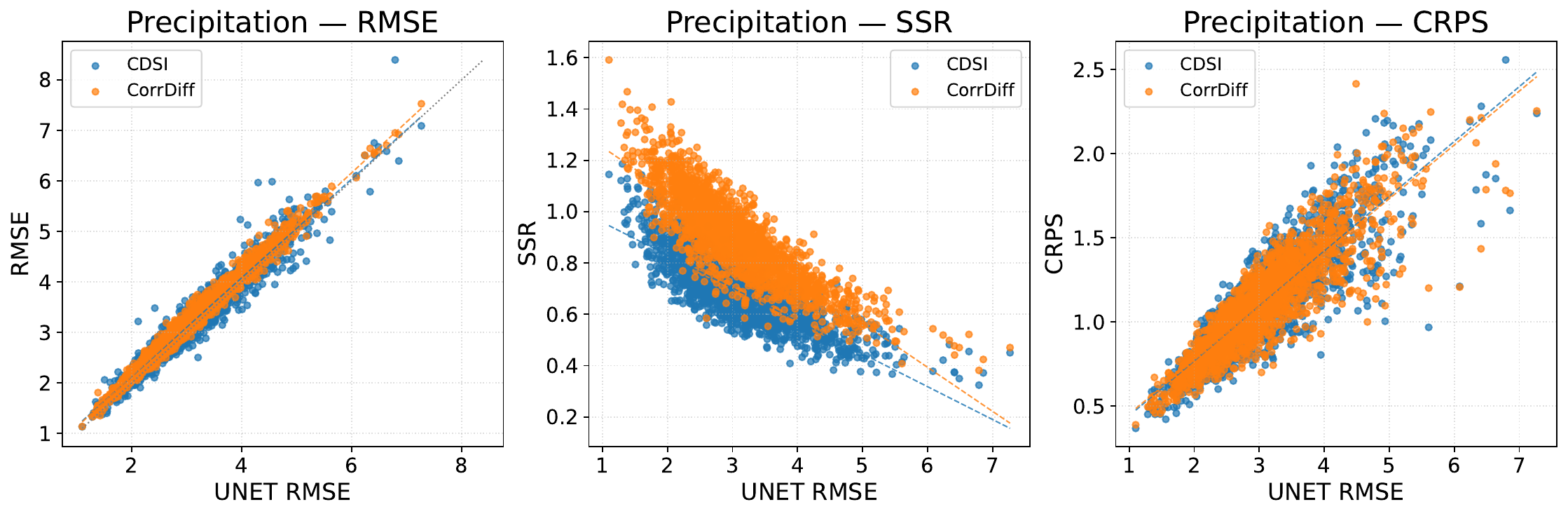}
    \caption{The RMSE, SSR, and CRPS as a function of the UNET RMSE for precipitation for the test period 2010--2014 for realization 1}
    \label{fig:metrics_vs_unet_pr}
\end{figure}

\begin{figure}[h!]
    \centering
    \includegraphics[width=\linewidth]{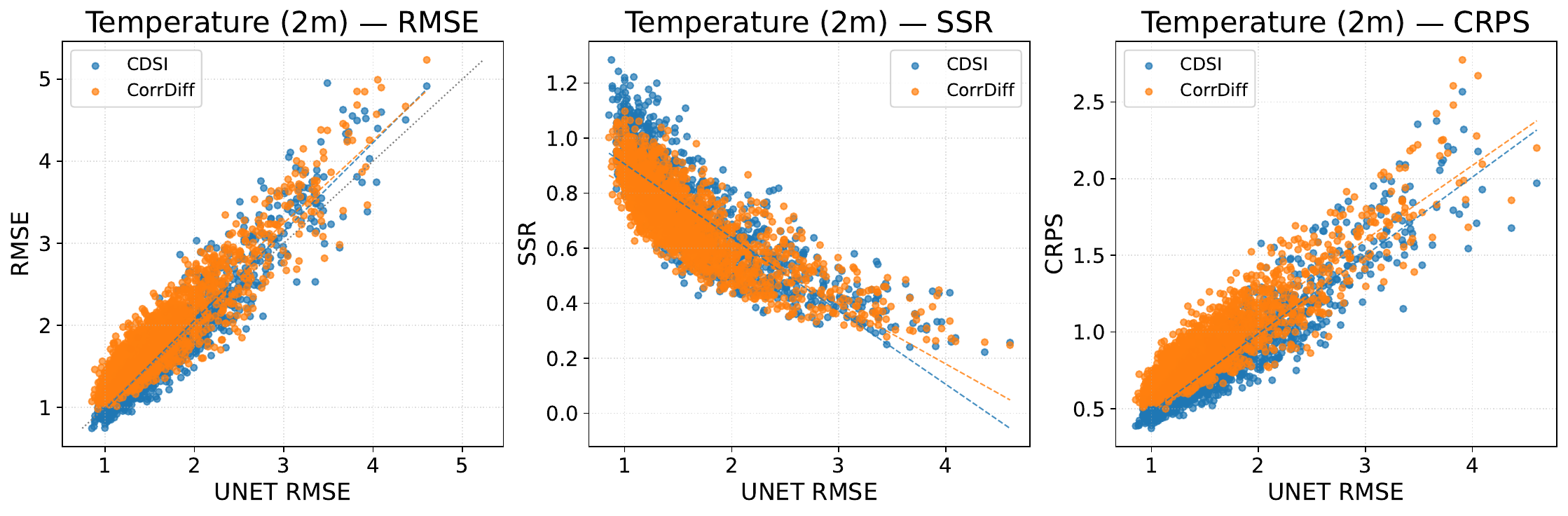}
    \caption{The RMSE, SSR, and CRPS as a function of the UNET RMSE for temperature for the test period 2010--2014 for realization 1}
    \label{fig:metrics_vs_unet_tas}
\end{figure}

Finally, \cref{fig:metrics_vs_unet_pr,fig:metrics_vs_unet_tas} relate the probabilistic metrics to the RMSE of the deterministic UNET baseline. This comparison provides insight into how probabilistic performance correlates with the difficulty of the underlying downscaling task. 

\end{document}